\documentclass[12pt]{article}
\usepackage{palatino,cite,graphicx,amsmath,amssymb,dcolumn}
\usepackage[bf,hang]{caption}
\usepackage{thumbpdf}
\usepackage[colorlinks=true, linkcolor=blue, citecolor=blue, urlcolor=blue,
pdfauthor={M.~Ball, N.~Ghodbane, M.~Janssen, P.~Wienemann},
pdftitle={A DAQ System for Linear Collider TPC Prototypes based on the ALEPH TPC Electronics},
pdfkeywords={TPC Readout}]{hyperref}

\begin{document}

\begin{flushright}
{\tt LC-DET-2004-013}
\end{flushright}
 
\vspace{3cm}
 
\begin{center}
    {{\Large \bf A DAQ System for Linear Collider TPC Prototypes\\[1ex]
                 based on the ALEPH TPC Electronics}\\[5ex]
    {\sc M.~Ball$^1$, N.~Ghodbane$^2$, M.~Janssen$^{1,3}$,
         P.~Wienemann$^1$}\\[3ex]}
    {\sl $^1$DESY, D-22607 Hamburg, Germany\\[1.5ex]
         $^2$CERN, CH-1211 Geneva 23, Switzerland\\[1.5ex]
         $^3$University of Dortmund, D-44221 Dortmund, Germany}

\end{center}
 
\vspace{3cm}
 
\begin{abstract}
\noindent
Within the international studies of a high energy linear electron positron 
collider, several groups are developing and testing prototypes for a Linear 
Collider TPC. This detector is planned to be used as a central part in the 
tracking system of a detector at such a machine. In this note we describe a 
DAQ system, which has been developed for the use in tests of 
TPC prototypes. It is based on electronics used at the ALEPH 
experiment at CERN. 
\end{abstract}

\thispagestyle{empty}

\newpage

\section{Introduction}

A large Time Projection Chamber (TPC) is proposed as part of the tracking
system for a detector at the future electron positron linear 
collider~\cite{ref:TeslaTDR}. To meet
the different milestones and requirements, several institutes have started a
joint effort to develop a technology based on the use of micro pattern
gas detectors like GEMs or Micromegas instead of the conventional wire
chamber based solution. A powerful but simple DAQ system is needed at 
this stage, based on already available technologies, to enable these
tests with minimum effort and expenditure. The DAQ system described in 
this note is based on the existing electronics used at the ALEPH experiment 
at LEP, CERN, for the readout of the ALEPH TPC. 

In this note we introduce the basic concept of the readout system, 
and discuss in detail each of the different components. 

\section{Principle of Operation}
The ALEPH TPC DAQ is based on an FADC system. Signals from each 
channel of the TPC are sent to a preamplifier, which integrates the 
charge and converts it into a voltage signal. The integration time 
constant can be influenced by a feedback capacitor, and is 
set in its default configuration to 2 $\mu$s. The signal 
from the preamplifier is sent to the digitizer via a twisted pair 
cable. The first stage of the digitizer is a receiver/shaper amplifier, 
followed by an 8-bit FADC, operated at 12.5 MHz. 
The digitized information is written 
into a memory bank, which can keep up to four different events of 
512 time slices each. The basic setup of the system is shown in 
Fig.~\ref{fig:daq}.

The required performance of a DAQ for the Linear Collider TPC has 
been summarized in the TESLA TDR \cite{ref:TeslaTDR}. The requested 
performance and the relevant parameters for the ALEPH 
system are summarized and compared in Tab.~\ref{tab-daq}.
\begin{table}[h]
  \begin{center}
    \begin{tabular}{l c c}
 Parameter      & TDR DAQ   & ALEPH DAQ \\ \hline
 Sampling Speed & $>$20 MHz & 12.5 MHz  \\
 ADC Range      & 9 bit     & 8 bit     \\
 Storage Depth  & 1 ms      & 512 time slices $\hat{=}$ 41 $\mu$s \\
\hline
    \end{tabular}
  \caption{\label{tab-daq}Table of requirements of the DAQ as specified in 
    the TESLA TDR and as delivered by the ALEPH based DAQ described in
    this note.}
\end{center}
\end{table}

\section{Hardware Setup}
The TPC DAQ is based on a combination of FASTBUS technology with VME based 
readout. The digitizer units (TPDs) are realised as FASTBUS modules,
through a slightly modified version of a standard FASTBUS crate is employed. 
Through a special link (see below for more details) the FASTBUS crate 
is controlled and read out from a VME based CPU. The complete system is 
controlled by an external computer, running the Linux operating system. 
In detail the following components are needed to assemble a complete system
(see Fig.~\ref{fig:daq}):
\begin{figure}
\begin{center}
    \includegraphics[width=0.9\textwidth]{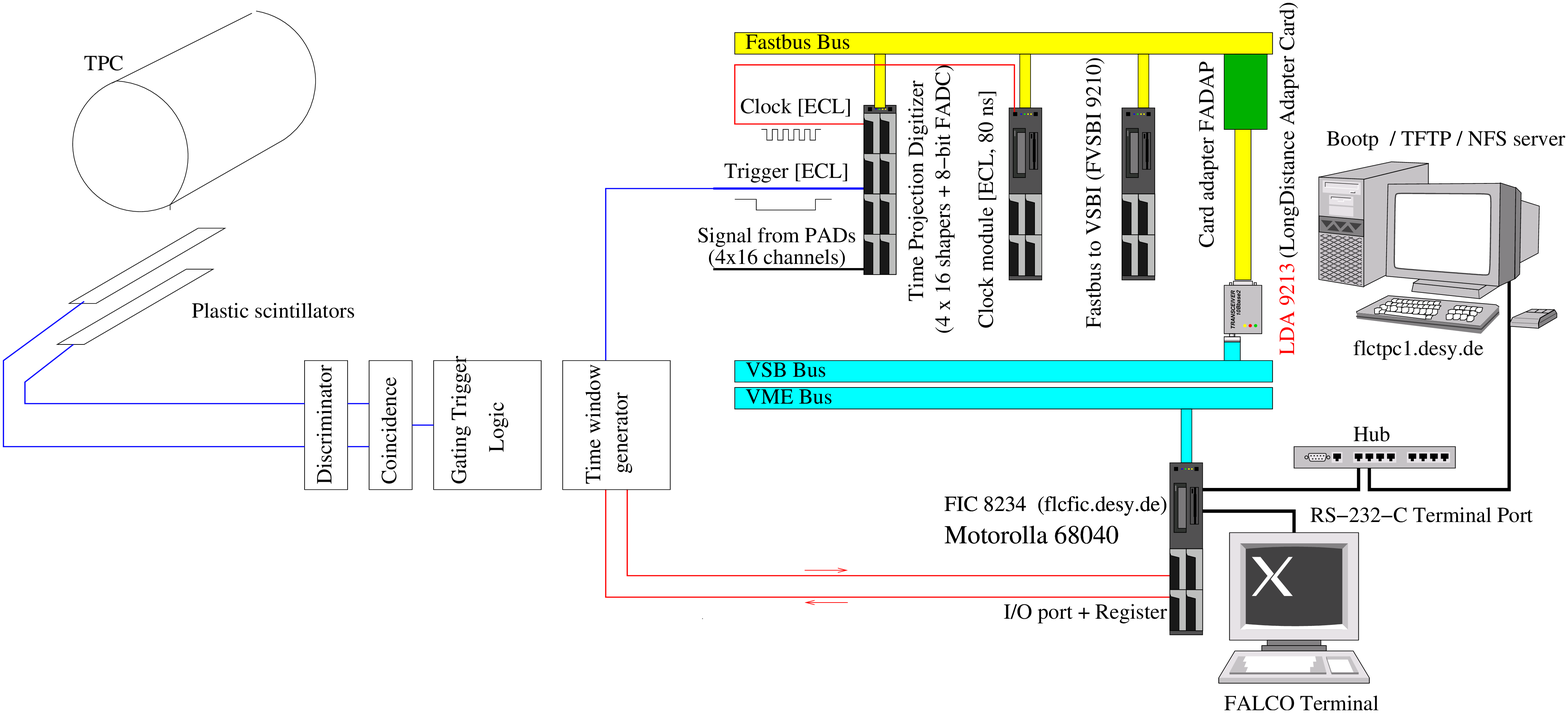}
\end{center}
\caption{Setup \`a la ALEPH for the TPC.}
\label{fig:daq}
\end{figure}

\begin{itemize}
\item One computer operating under Linux (SuSE 7.3 operating system),
\item A 6U VME crate with:
\begin{itemize}
\item A Fast Intelligent Controller (FIC 8234~\cite{ref:FIC8234}),
      A 68040 VME/VSB dual processor unit running at 25 MHz. 
\item A VME/VSB adapter card, Long Distance Adapter LDA 9212 or LDA 9213.
\end{itemize}
\item A modified FASTBUS crate with:
\begin{itemize}
\item A FASTBUS to VSB translator unit (FVSBI 9210~\cite{ref:FVSBI9210}) 
\item A set of Time Projection Digitizer (TPD 6821~\cite{ref:TPD6821}) modules
\item One FASTBUS Adapter Card, FADAP 9211.
\end{itemize}
\item A NIM crate containing several electronic modules for the trigger
  selection. The trigger signal has to be converted finally into an ECL signal,
which can be used by the TPD.
\end{itemize}
In the following sections the way of the signal through the
electronic chain is described. A complete description of the ALEPH TPC
electronics can be found in \cite{ref:AlephHandbook}.

\subsection{The Preamplifier and Shaping Amplifier}
Conventional TPCs like that of ALEPH have used a multi-wire
proportional chamber technique to multiply the primary
electrons at the endplates. With this method, thin anode wires are
mounted just in front of the readout pads. Due to the high electric
field in the vicinity of the wires, electron multipication takes place
so that the produced charges induce measurable signals on the readout
pads. Unfortunately this technique leads to a dependence of the
resolution on the projected angle between track and wires. Moreover
the electric and magnetic field lines are not parallel close to the
wires leading to a significant drift velocity component along the
$\vec{E} \times \vec{B}$ direction.  This eventually might limit the
spatial resolution of the chamber. In addition wires and the necessary
support structures add large amounts of material to the endplate, and
thus compromize the performance of the calorimetry in the forward
direction.

For the Linear Collider therefore a novel readout scheme for a TPC is
proposed, based on micro-pattern gaseous chambers. In recent years
GEMs and Micromegas have been developed to a point where their use in
large scale detectors can be envisioned.  Contrary to the wire chamber
readout, TPCs with a GEM or Micromegas based gas amplification system
directly detect electrons produced in the amplification step. Since
this signal is very fast, the parameters of the preamplifiers need to
be specially optimized for this situation.

The input stage of the ALEPH preamplifier is a charge-integrating
circuit with a decay time constant of 2 $\mu$s. The effective feedback
capacitor is $C_f$ = 1 pF. The charge sensitivity is determined by the
feedback capacitor $C_f$ to be 1 V/pC. For a single pad of size 2
$\times$ 6 mm a typical primary electron signal is around 25
electrons.  After gas amplification this translates into a
preamplifier output signal of $\frac{f \cdot g \cdot
  q}{C_f}\approx$ 12 mV with the charge collection efficiency $f$
due to the integration time, the gas gain $g$, the primary ionization
charge $q$ and the capacitance $C_f$ of the storage capacitor.

The output signal of the preamplifier rises very fast -- typical rise
times are of the order of a few tenth of a ns -- and then decays
exponentially. The shaper symmetrizes the signal and converts it into
a roughly Gaussian shape with a width of about 400 ns. The FADC then
digitizes this signal at 12.5 MHz, resulting in a typically length of
5 time slice samples for one pulse.

Technically the preamplifiers are connected to the TPDs via twisted pair 
cables. The preamplifiers are supplied with power through the 
signal cables. $\pm$ 5 V needs to be supplied either from an 
external power supply or through a modification of the FASTBUS crate. 

Both the preamplifiers and the shaping amplifiers were produced in
thick-film hybrid technology.  Measurements done at the time of
production of the electronics showed that the gain and the noise of
the devices were constant within an RMS of 1.5 \% for the
preamplifier gain, $\pm$ 5 ns for the FWHM of the output pulses after
shaping.

A special FASTBUS module (SMTPD) exists which, apart from other things, 
provides the necessary 5 V to a modified FASTBUS backplane, so 
that the supply voltage can be applied to the preamplifiers.

\subsection{The TPD}
\label{sec:TPD}

In this Section we discuss the Time Projection Digitizer (TPD). A more
complete description can be found in~\cite{ref:TPD6821}.\\

The TPDs are implemented in FASTBUS technology and adhere to the 
FASTBUS standard. The modules are controlled through the 
FASTBUS control bus. 

The TPD, originally designed to digitize the signals of the
ALEPH TPC, comprises 4 $\times$ 16 input channels, each of them built
around an 8-bit Flash Analog To Digital Converter (FADC) preceded by a
hybridized line receiver/shaper amplifier. The TPD uses the standard ALEPH TPC
preamplifiers as signal sources.

The TPD supplies power to the preamplifiers as described in the 
previous Section. For testing purposes a specially designed module 
is available (SMTPD). This module in addition provides clock,
trigger and pulse signals and is primarily used during the 
testing and debugging phase of the electronics. 

Later during routine operation an external source was used to 
supply the $\pm$ 5 V to the preamplifiers. 

When using old ALEPH equipment special care needs to be taken 
to make sure that a properly modified FASTBUS crate is 
available when the SMTPD modules are used. 

\subsubsection{Front Panel Input Signals Handling}
A TPD accepts three different kinds of input signals:
\begin{itemize}
\item The clock (CK) at a frequency of 12.5 MHz, used to sample the
  signal from the preamplifiers. The clock has to be a standard NIM
  signal.
  
  The clock can be either supplied from the SMTPD -- which is suitable
  for a small number of TPDs -- or through an external clock
  generator.  Simple ways of realizing such an external clock
  generator are given in the Appendix~\ref{sec:ClockHowto} to this note.
  
\item The trigger or write signal (WR). The WR signal starts the
  acquisition procedure of the FADCs. The WR signal duration must be
  greater than 512 time slices $\times$ the CK period, i.~e.~41 $\mu$s.
  
\item The signals from the preamplifiers connected to the pads of the
  TPC.
\end{itemize}

Once the TPD gets the WR signal, each of the 64 input signals is passed
through the shaper and the FADC, with a typical sampling time
of 80 ns (CK signal). The digitized information is piped into a 
buffer which is 512 bytes deep, called the Raw Data Memory (RDM). Four 
parallel banks are available, allowing to store up to 4 events in 
the system before the events have to be readout into the 
acquisition computer. 

The events stored in the TPD are eventually written into static memory, 
after applying some data selection criteria like threshold suppression, 
pulse time extraction, etc. 

\subsubsection{Control Status Registers}

The full functionality of the TPD is available through 
various internal 32 bit registers. 
These are written and read by the FIC 8234 via the VSB to FASTBUS interface
FVSBI 9210. 

The TPD implements several control and status registers (CSR). 
The most relevant ones, which are used in the acquisition system are:
\begin{itemize}
\item \texttt{CSR\#0}: This register is a general purpose control and status
  register implemented according to the FASTBUS specifications. It
  contains the TPD module identifier and the different control bits
  for proper operations.  The \texttt{CSR\#0} bit significance is given in
  Tab.~\ref{tab:CSR0}.
\item \texttt{CSR\#1} is used for the access and DAQ Bank
  informations.  The \texttt{CSR\#1} bit significance is given in
  Tab.~\ref{tab:CSR1}.
\item Two Next Transfer Address (NTA) registers, one for the DATA
  space and one for the CSR space.
\item \texttt{CSR\#0xC0000002} or DAC register implements four 6-bit values, so
  called DAC values to set the parameters of each FADC
  channel. These four values control the pedestal, linearity and the
  gain of the TPD.
\end{itemize}

\subsection{The FASTBUS to VME/VSB Interface (FVSBI)}
\label{sec:FVSBI}

The FVSBI 9210 connects the VME to the FASTBUS, and thus allows the
control of the FASTBUS system.  It communicates with the VSB bus using
two extension cards, a Long Distance Adapter card (LDA 9213) which is
plugged into the rear of the VSB connectors. A passive adapter, the
F-side Adapter (FADAP 9211), is connected at the rear of the FASTBUS
crate to the LDA 9212.

There are two directions of communication.  The FIC can control and
set all relevant registers etc.~of the TPDs in the FASTBUS crate.
After the digitization step, the FIC reads the memory in the TPDs and
transfers the data through the VME crate to the DAQ computer.


\section{Acquisition Software Description}

\subsection{Approach Using the RDM Banks}

Each of the 64 8-bit FADCs digitizes the signal from the ALEPH preamplifiers
and stores this event in the so called Raw Data Memory (RDM). A simple
approach consists in reading this memory once the digitization is done and
apply offline threshold cuts and sophisticated algorithms to find the
different charge clusters.

This simple approach has been implemented in the acquisition software and the
different steps are:

\begin{itemize}
\item Initialization of the FASTBUS session.
\item Initialization of the FVSBI master (see Section~\ref{sec:FVSBI}).
\item Loop on the different FASTBUS slots and for each primary address (PA)
 try  to read the CSR\#0, where the identity of the FASTBUS module is written
 and store the PA associated with the different modules (TPDs and SMTPDs).
\end{itemize}

Having the list of TPDs in use, then, for each of them:
\begin{itemize}
\item Reset the TPD and all CSRs by setting the \texttt{CSR\#0<30>} to 1.
\item Set the DAQ bank to be bank 0, 1, 2 or 3 using \texttt{CSR\#1<19:18>}.
\item Load the DAC values used by the 8-bit FADCs.
\item Enable DAC writing by setting \texttt{CSR\#0<12>}.
\item Loop on the 64 channels and for each of them:
\begin{itemize}
\item Define the channel to be loaded in the NTA\_DATA register \texttt{NTA\_DATA<14:09>}.
\item Load the DAC value in the \texttt{CSR\#0xC0000002} register.
\item Wait for bit clear of \texttt{CSR\#0<13>}.
\item Wait a bit that the DAC value loaded is stable.
\end{itemize}
\item Disable the DAC serializing, setting \texttt{CSR\#0<28>}.
\item Write to the Least Significant Bit (LSB) of CSR\#1, \texttt{CSR\#1<00>}.
\item Enable the DAQ, \texttt{CSR\#0<06>}.
\item If a new trigger is accepted, then a WR cycle starts and
  \texttt{CSR\#0<07>} goes up to 1. Then an image of the current number of time
  samples  can be found in the DAQ counter \texttt{CSR\#1<07:00>}.
\item The DAQ stops if the DAQ counter reaches 511 and then the
  \texttt{CSR\#0<07>} is reset to 0.
\item Disable the DAQ setting \texttt{CSR\#<22>}.
\item Set the Access Bank to read to be bank 0, 1, 2 or 3 using \texttt{CSR\#1<17:16>}.
\item Read the data channel by channel or block by block.
\end{itemize}

\subsection{Acquisition Software}

The TPD readout has been described in detail in the Section before. The
{\texttt{C }}  program {\texttt{tpcdaq.c}} is well commented and the names of the
different implemented functions are self-explanatory. In this
Section, we list the different implemented functions and summarize briefly
their task. These functions are called from the main program {\texttt{TPCAcquire()}} and are:
\begin{itemize}
\item {\texttt{FastBusInit()}}: Initializes the FASTBUS session.
\item {\texttt{FastBusScan()}}: Scans the different FASTBUS segments  and
 returns the PA of the different TPDs and SMTPDs present.
\item {\texttt{getFVSBInterfaceNumber()}}: Returns the FVSBI 9210 interface
  number set on the front panel. 
\item  {\texttt{getFastbusModuleID()}}: Tries to read the CSR\#0 at a given PA and
  decode the module identity coded in this register.
\item  {\texttt{FastBusClose()}}: Finishes the FASTBUS session.
\item  {\texttt{checkFBError()}}: Decodes the FASTBUS error if any.
\end{itemize}

The SMTPD is not crucial for the data acquisition since it can easily be
replaced by a NIM clock module. 
Nevertheless, as described above, it enables to test and calibrate the
TPD. Thus several functions have been implemented starting from the definition of
the different registers in reference~\cite{ref:SMTPD6917}. These functions are:

\begin{itemize}
\item  {\texttt{SMTPDReset()}}: Reset the SMTPD.
\item  {\texttt{SMTPDClock()}}: The clock is not automatically started at the
  power up of the SMTPD. this has to be done by software.
\item  {\texttt{SMTPDSetSource()}}: The SMTPD can be used as a testing module
  which generates the input signals for the TPD. 
This function has not been tested.
\end{itemize}

To access the different TPDs several functions have been implemented. 
These functions are:
\begin{itemize}
\item  {\texttt{TPDReset()}}: Reset the TPD module.
\item  {\texttt{TPDSetDAQBank()}}: Select one of the four banks to which the sampled event will  be written to.
\item  {\texttt{TPDSetDAC()}}: Set the DAC parameters to control the slopes of the
  different FADC channels.
\item  {\texttt{TPDEnableDAQ()}}: Prepare the TPD for the acquisition of the next event.
\item  {\texttt{TPDSetDAQ()}}: Prepare the TPD for the acquisition setting \texttt{CSR\#0<07>}.
\item  {\texttt{TPDDAQActive()}}: Checks whether the  \texttt{CSR\#0<07>} set previously
  is still up. At the end of the digitization procedure this bit returns to
  0. 

\item  {\texttt{TPDWaitForTrigger()}}: Reads the CSR\#1 time slice counter set
  initially to 1 and incremented during the digitization of the pulse. At the
  returns, this reaches 512 time samples. 

\item {\texttt{TPDDisableDAQ()}}: Disable DAQ setting \texttt{CSR\#0<22>}.
\item {\texttt{TPDSetAccessBank()}}: Set the access bank to one of the four
  banks, where the data have been stored during digitization.
\item {\texttt{TPDReadFADCCharge()}}: Read the bank where the raw data are stored.
\item {\texttt{TPDReadRDM()}}: Read the event from the Raw Data Memory (RDM)
  directly. As explained above. No selection criteria is applied to the event in the  TPD.
\end{itemize}

The TPD is capable of applying selection criteria to the raw data
stored in the RDM. To do this, several functions have been implemented.
These functions are\footnote{At the time of writing this note, this part
of the system has not yet been extensively tested. Please contact the
authors to find out about the current state of this part of you intend
to use these functions}:
\begin{itemize}
\item  {\texttt{TPDSetThreshold()}}:  Set the threshold to be used.
\item  {\texttt{TPDSetLimitRegister()}}: Set the Limit register to define the
  cluster (total number of time slices and number of time slices below threshold).
\item  {\texttt{TPDDolist()}}: Start or apply the selection on the data stored in the TPD banks.
\item  {\texttt{TPDReadHLM()}}: Read the data stored in the Hit List Memory (HLM)
  after the selection.
\end{itemize}

The FIC 8234 implements a front panel Input/Output controller, which can be
used to generate e.~g.~a gate (e.~g.~with a FLIP/FLOP). Several functions are provided to drive
this controller. These are {\texttt{FPInit()}, \texttt{FPClose()} and {\texttt{FPWrite()}}.\\

The data read from the TPD by the FIC 8234 can either be stored 
using NFS or 
using the more powerful client/server approach explained  below. The drawback
of the first approach is that the FIC 8234 communicates only with one
computer, its NFS server, whereas the second approach enables the FIC to
communicate with several computers which can run reconstruction programs in
parallel. Several {\texttt{C}} functions have been implemented for this
purpose. These are {\texttt{NFSFileConnect()}},
  \texttt{NFSFileDataSend()} and {\texttt{NFSFileDisconnect()}} for the approach using the
NFS transfer protocol and {\texttt{TCPIPClientDataSend()}} for the TCP/IP client
server based solution.

\section{Network and Server Configuration}
The master processor of the whole data acquisition system is the Fast
Intelligent Controller, the FIC 8234.  The FIC is operated in
diskless mode. This means that upon startup the operating system has
to be loaded through the network from a network disk drive.

The FIC uses the BOOTP and TFTP protocols to find and upload its operating system: 
\begin{enumerate}
\item The FIC broadcasts a BOOTP request to find a BOOTP server.
\item If a BOOTP server is present it replies to the BOOTP request and
  sends the network settings (IP address, gateway address, name server
  address, TFTP server address, etc.) to the FIC.  Care has to be
  taken that a proper EPROM is used with an up-to-date
  version\footnote{Older EPROMs send BOOTP requests with
    255.255.255.255 as source address. Such packets are ignored by the
    BOOTP server. The correct source address is 0.0.0.0. You have to
    ask CES Electronics to replace OS-9/68040 2V4 Rev. A2 by
    OS-9/68040 2V4 Rev. 1.4.}.
\item The FIC reboots itself with the settings it has received from
  the BOOTP server.
\item The next step of the network configuration is that the FIC
  starts a request for a TFTP server to load the boot image of the
  OS-9 operating system.
\item The last step of the configuration is to establish a NFS
  connection to the Linux PC to have access to the file space where
  the acquisition software is located.
\end{enumerate}   
The different steps to implement all these technical
modifications are presented on the web page
\href{http://www.cern.ch/ghodbane/tpc}{\tt http://www.cern.ch/ghodbane/tpc}.

The system has been developed using Linux, SuSE 7.3 professional 
edition, with the following software versions:
\begin{itemize}
\item Linux Kernel: 2.4.10-4GB.
\item BOOTP server: bootp-DD2-4.3-87 (as an RPM).
\item TFTP server: tftp-0.20-22 (as an RPM).
\item Kernel NFS server and NFS utilities nfs-utils-0.3.1-87 (as an RPM).
\end{itemize}
Newer distributions should work as well, but have not been tested. 




\section{Client Server Based Data Distribution}

The data processed by the FIC CPU are sent via a TCP connection to a central
data distribution server (DDS) running on a Linux PC. This server forwards all
data received from the FIC to any other clients connected to it.  Possible
data receiving clients are data writing programs to make the data persistent,
online monitoring software to check the data quality, etc. In principle an
arbitrary number of clients can connect to the DDS. In order to minimize the
load of the FIC CPU and to keep the data transfer through the slow 10 MBit/s
connection of the FIC as low as possible, the server does not run directly on
the FIC but on a separate machine. A sketch of the setup is shown in
Fig.~\ref{fig:clientserver}.

\begin{figure}
\begin{center}
    \includegraphics[width=0.8\textwidth]{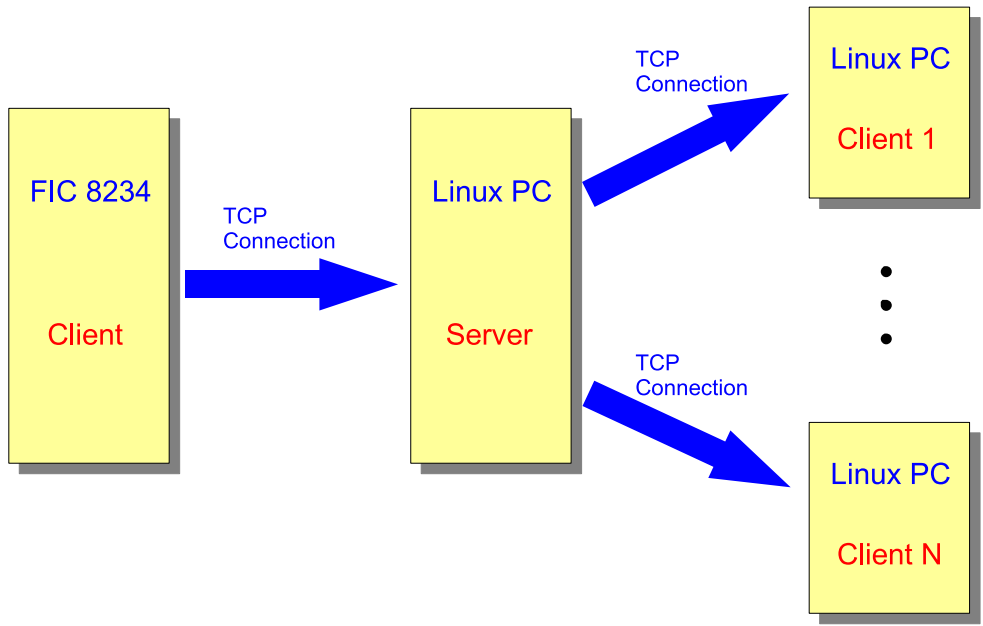}
\end{center}
\caption{The client server based data distribution system used by the DAQ
  system. Data processed by the FIC CPU are sent to a central data
  distribution server which in turn forwards the data to an arbitrary number of
  clients connected to it.}
\label{fig:clientserver}
\end{figure}

\subsection{Data Transfer Protocol}

The data are sent in chunks as 8 bit {\tt char} arrays to the DDS. Since all
numbers to be transferred are integers, they are represented either as
8 bit {\tt char}s or 32 bit {\tt int}s, depending on their range.
An {\tt int} is packed into a {\tt char} array by distributing the 32
{\tt int} bits over multiple {\tt char}s.

At the beginning of each event, a {\tt 0x61='a'} character is sent to
indicate the start of a new event. This is followed by an event number, the
UNIX time when the event has been recorded and the number of TPDs used. These
numbers are filled sequentially into a character array which is transferred to
the DDS {\it en bloc}. Then for each TPD, the TPD ID and the number of pulses
is filled into a character array and sent to the DDS. The meaning of "pulse"
depends on the user requirements. It can be just a couple of time slices in a
channel which exceed a certain threshold value, but it can also be the full
FADC spectrum of a channel. The latter definition has been adopted as the
default setting, i.~e.~the number of pulses is equal to the number of
channels. The next data to be transferred in a large array is the contents of
each time slice for all the pulses. Finally at the end of each event, a {\tt
  0x65='e'} character is sent to indicate the end of the event.

In order to avoid control characters from being transferred, the
protocol works on a 7-bit per character basis. The most significant
bit (MSB) of each character is always set to one\footnote{Note that
  all control characters have a MSB set to zero.}, so that only 7
payload bits remain per character. The only exceptions with MSB off
are {\tt 0x61='a'} and {\tt 0x65='e'} which indicate the beginning and
the end of an event, respectively.

\subsection{LCIO Data Writing Client}

The developed setup uses LCIO \cite{ref:LCIO}, the Linear Collider
data model and persistency framework.  A client has been implemented
which connects to the DDS and writes the transmitted data to disk in
LCIO format.

The LCIO classes are kept as general as possible in order to avoid any
unneeded restrictions. The class {\tt IMPL::TPCHitImpl} provided by LCIO is
designed in such a way that data from both FADC and TDC based TPC readouts can
be stored without unnecessary overhead. Therefore a FADC specific class called
{\tt TPCPulse} has been added to the LCIO class collection to allow easy
access to all relevant information. This class is defined in the following way:
{\footnotesize
\begin{verbatim}
#ifndef LCIO_TPCPULSE_H
#define LCIO_TPCPULSE_H 1

#include "LCObjectHandle.h"
#include "IMPL/TPCHitImpl.h"


typedef lcio::LCObjectHandle< IMPL::TPCHitImpl > TPCHitImplHandle ;


class TPCPulse : public TPCHitImplHandle {
public:
    TPCPulse( DATA::LCObject *lcObj) : TPCHitImplHandle(lcObj)  {;}
    TPCPulse( IMPL::TPCHitImpl *lcObj) : TPCHitImplHandle(lcObj)  {;}

    virtual ~TPCPulse() {;}

    int     getNBins() const         { return _lcObj->getNRawDataWords() - 1; }
    int     getTime(int bin) const   { return _lcObj->getRawDataWord(0) + bin; }
    int     getCharge(int bin) const { return _lcObj->getRawDataWord(bin + 1); }
    int     getTPDID() const         { return _lcObj->getCellID() / 100; }
    int     getChannel() const       { return _lcObj->getCellID() % 100; }

    void    setPulseSpectrum(int tpd, int channel, int time0,
                             const int* charge, int nbins){
        lcObj()->setCellID(tpd * 100 + channel);
        int* a = new int[nbins+1];
        a[0] = time0;
        for (int i=0; i<nbins; i++) a[i+1] = charge[i];
        lcObj()->setRawData(a, nbins+1);
        delete[] a;
    }

};

#endif
\end{verbatim}
}

\begin{figure}
\begin{center}
    \includegraphics[width=0.8\textwidth]{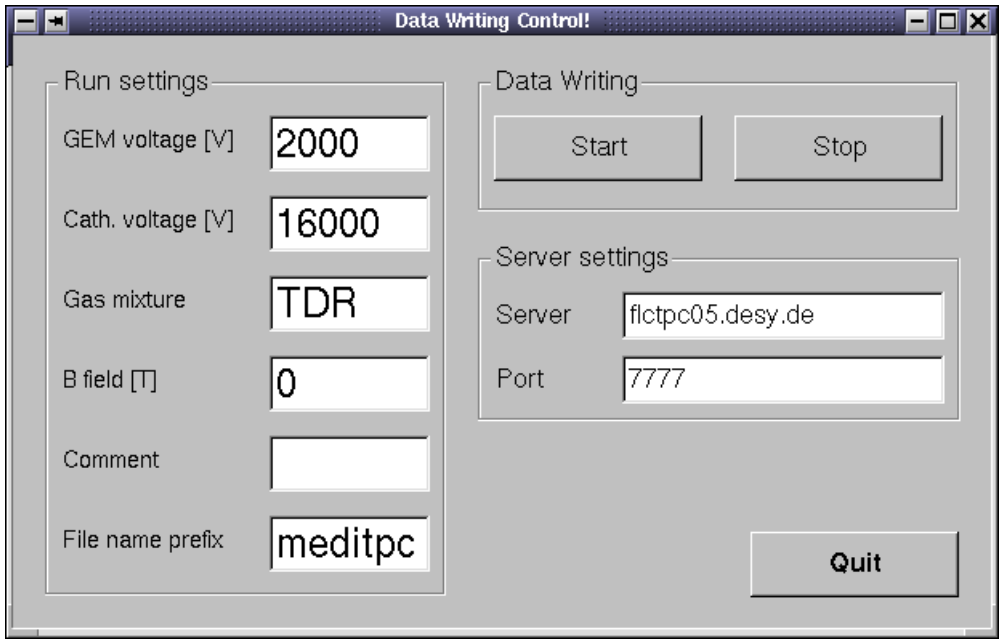}
\end{center}
\caption{A screen shot of the control window of the LCIO Data Writing Client.}
\label{fig:lcioclient}
\end{figure}

A screen shot of the program control window is shown in
Fig.~\ref{fig:lcioclient}. In addition to the data from the readout
electronics, important run parameters like cathode voltage, magnetic field
value etc.~can be given. They are stored in the run description
string\footnote{Each field is separated by a "$|$". Therefore a "$|$" should
  not be contained in the entered string to avoid confusion.} available in
each LCIO run header. The file name prefix provided by the user is used to
compose the data file name.  It has the following components: {\tt
  prefix.YYYYMMDD.XXX.slcio}. {\tt YYYYMMDD} is the date when the run has been
started and {\tt XXX} a serial number starting from {\tt 000}. Every 1000
events a new file is created and its serial number is increased by one.

Data recording is started by pressing the start button. Only after the start
button has been pressed, an LCIO file is opened and a TCP connection to the
given server and port is established. A run can be stopped by pushing the stop
button which closes the LCIO data file and the TCP connection to the DDS.

\section{Summary and Conclusion}
A DAQ has been developed which is based on existing hardware which was 
used in the ALEPH experiment at LEP. The DAQ offers a low-cost 
solution to the need of having DAQ systems of intermediate power 
available during the R\&D phase for a LC TPC. The DAQ offers 
reasonable performance, and can be used to study setups of 
reasonable complexity. 

It is clear however that this DAQ system can not be considered 
as a prototype for the eventual acquisition system a the LC TPC. 
Acquisition speed, package density are too low, the power 
consumption is much too high for this. Significant R\&D is needed 
to develop and design a powerful yet compact readout system 
for a future LC TPC. 

\section{Acknowledgments}
We are grateful to B.~Jost, B.~Lofstedt and R.~Schuler for their
valuable help.  We also wish to express our gratitude to F.~Gaede.
Working closely together made the rapid inclusion of the TPCHit
classes into LCIO possible. In addition we would like to thank
T.~Behnke for reading the manuscript and providing fruitful comments.
We strongly encourage the R\&D groups to systematically write such a
note where they clearly describe how they have built the setups.

\begin{appendix}

\section{Building a Simple External Clock Generator}
\label{sec:ClockHowto}

A simple external clock can be built using a dual gate generator. The
setup described in this note has been tested with a LeCroy NIM Module,
model 222. The required cabling for such a clock is shown in
Fig.~\ref{fig:clock}.

\begin{figure}
\begin{center}
    \includegraphics[width=0.9\textwidth]{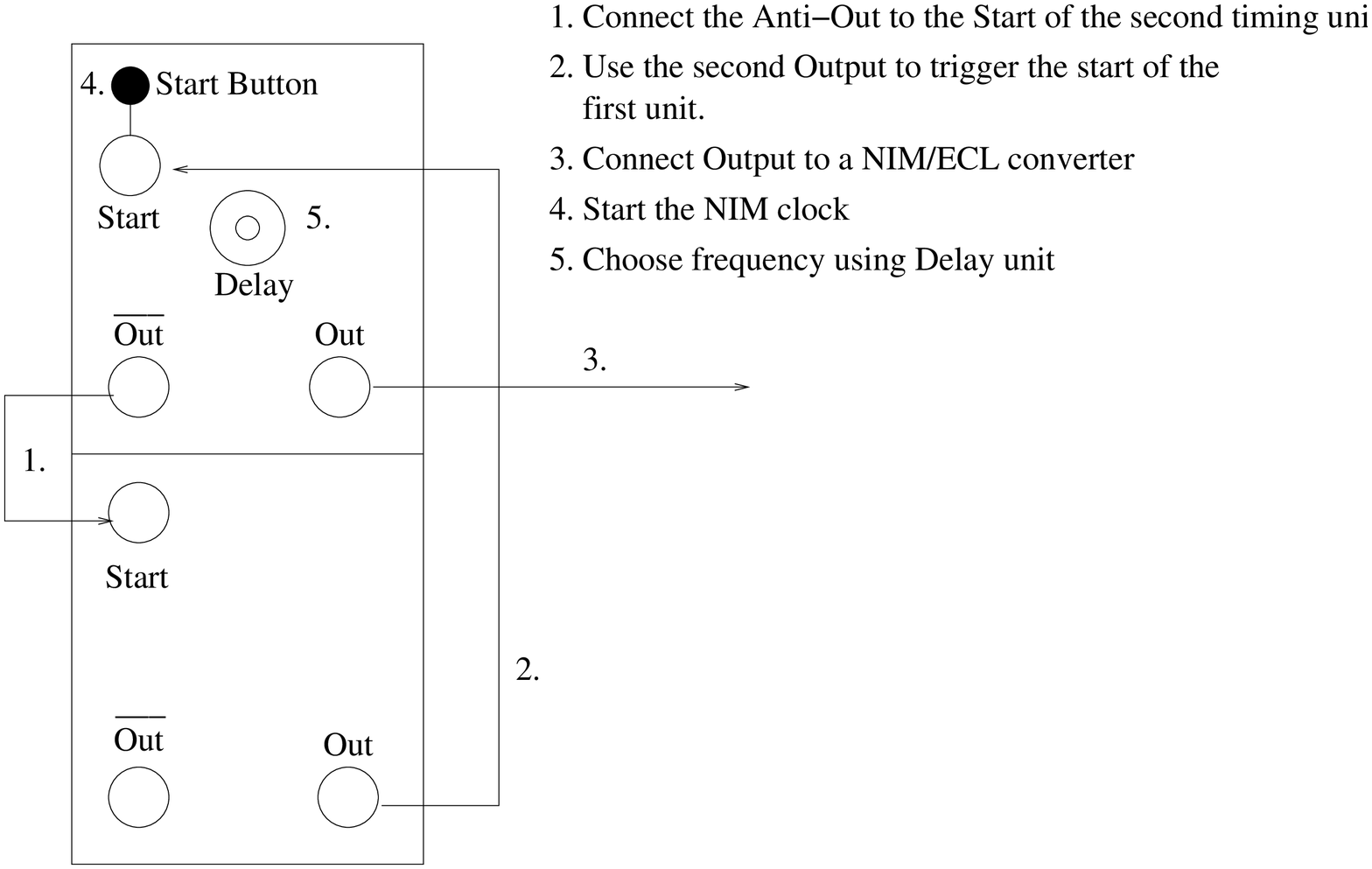}
\end{center}
\caption{Setup of an external clock generator using a dual gate generator.}
\label{fig:clock}
\end{figure}

\newpage

\begin{table}
\begin{center}
\begin{tabular}{c l l}
\hline
Bit & Read Significance & Write Significance \\
\hline
00 & Error flag        & set error flag\\
04 & Serialize (SR) assertion enabled & Enable SR assertion \\
05 & SR asserted &\\
06 & DAQ enabled & Enable DAQ \\
07 & Status DAQ active& \\
08 & Status Dolist active & enable Dolist\\ 
09 & Status hit counter overflow & \\
11 & Test input enabled & \\
12 & Digital to Analog Converter (DAC) enabled & Enable DAC writing \\ 
13 & Status DAC serial load &\\
20 & & Disable SR assertion\\
21 & & Reset serialize (SR)\\
22 & & Diable DAQ\\
24 & & Disable Dolist\\
28 & & Disable TEST input\\
29 & & Disable DAC serial load\\
30 & & Reset module (including CSR) \\
31 & & Reset registers and counter but not CSR\\
\hline
\end{tabular}
\caption{CSR\#0 bit assignement}
\label{tab:CSR0}
\end{center}
\end{table}

\begin{table}
\begin{center}
\begin{tabular}{c l}
\hline
Bit & Significance \\
\hline
08:00 & DAQ counter \\
17:16& Access bank\\
19:18& DAQ bank\\
\hline
\end{tabular}
\caption{CSR\#1 bit assignement}
\label{tab:CSR1}
\end{center}
\end{table}

\begin{table}
\begin{center}
\begin{tabular}{c l l}
\hline
Bit & Read Significance & Write Significance\\
\hline
30    & & Reset the SMTPD\\
31:16 & Module identifier: 0x6917 &\\
\hline
\end{tabular}
\caption{CSR\#0 bit assignment for the SMTPD}
\label{tab:SMTPDCSR0}
\end{center}
\end{table}

\begin{table}
\begin{center}
\begin{tabular}{c l l}
\hline
Bit & Read Significance & Write Significance\\
\hline
07:00& & Pulse amplitude value\\
\hline
\end{tabular}
\caption{CSR\#1 bit assignment for the SMTPD}
\label{tab:SMTPDCSR1}
\end{center}
\end{table}

\begin{table}
\begin{center}
\begin{tabular}{c l l}
\hline
Bit & Read Significance & Write Significance\\
\hline
06& CP out enabled & Enable CP generation\\
07& Signal pulse enabled & Enable signal pulse\\
08& Pulse routed to channel output & Route pulse to channel output \\
13& CP internaly generated & Enable internal CP oscillator \\
\hline
\end{tabular}
\caption{CSR\#2 bit assignment for the SMTPD}
\label{tab:SMTPDCSR2}
\end{center}
\end{table}

\end{appendix}

\end{document}